\documentclass[aps,prx,twocolumn,groupedaddress,showpacs,floatfix]{revtex4}

\usepackage{graphicx}
\usepackage{grffile}
\usepackage{subfigure}
\usepackage{amsmath}
\usepackage{amssymb}
\usepackage{bm}
\usepackage{nicefrac}
\usepackage{color}
\usepackage[dvipsnames]{xcolor}
\graphicspath{{./}{figures/}}

\newcommand{\etcucl}{$\kappa$-(ET)$_2$Cu[N(CN)$_2$]Cl}
\newcommand{\etcubr}{$\kappa$-(ET)$_2$Cu[N(CN)$_2$]Br}

\newcommand{\brcl}{$\kappa$-(ET)$_2$Cu[N(CN)$_2$]Br$_x$Cl$_{1-x}$}
\newcommand{\Epa}{\mbox{$\mathbf{E}\parallel\mathbf{a}$}}
\newcommand{\Epc}{\mbox{$\mathbf{E}\parallel\mathbf{c}$}}

\newcommand{\ie}{\it{i.\,e.}}

\begin{document}

\title{Unveiling the microscopic nature of correlated organic
conductors: the case  of {\brcl}}

\author{Johannes Ferber}
\affiliation{Institut f\"ur Theoretische Physik, Goethe-Universit\"at Frankfurt,
Max-von-Laue-Strasse 1, 60438 Frankfurt/Main, Germany}

\author{Kateryna Foyevtsova}
\affiliation{Institut f\"ur Theoretische Physik, Goethe-Universit\"at Frankfurt,
Max-von-Laue-Strasse 1, 60438 Frankfurt/Main, Germany}

\author{Harald O. Jeschke}
 \email{jeschke@itp.uni-frankfurt.de}
\affiliation{Institut f\"ur Theoretische Physik, Goethe-Universit\"at Frankfurt,
Max-von-Laue-Strasse 1, 60438 Frankfurt/Main, Germany}

\author{Roser Valent\'\i}
\email{valenti@itp.uni-frankfurt.de}
\affiliation{Institut f\"ur Theoretische Physik, Goethe-Universit\"at Frankfurt,
Max-von-Laue-Strasse 1, 60438 Frankfurt/Main, Germany}

\date{\today}

\begin{abstract}

  A few organic conductors show a diversity of exciting properties
  like Mott insulating behaviour, spin liquid, antiferromagnetism, bad
  metal or unconventional superconductivity controlled by small
  changes in temperature, pressure or chemical substitution.  While
  such a behaviour can be technologically relevant for functional
  switches, a full understanding of its microscopic origin is still
  lacking and poses a challenge in condensed matter physics since
  these phases may be a manifestation of electronic correlation.  Here
  we determine from first principles the microscopic nature of the
  electronic phases in the family of organic systems {\brcl} by a
  combination of density functional theory calculations and the
  dynamical mean field theory approach in a new form adapted for
  organic systems.  By computing spectral and optical properties we
  are able to disentangle the origin of the various optical
  transitions in these materials and prove that correlations are
  responsible for relevant features.  Remarkably, while some
  transitions are inherently affected by correlations, others are
  completely uncorrelated. We discuss the consequences of our findings
  for the phase diagram in these materials.

\end{abstract}

\maketitle

\begin{figure*}[htb]
\includegraphics[width=0.75\textwidth]{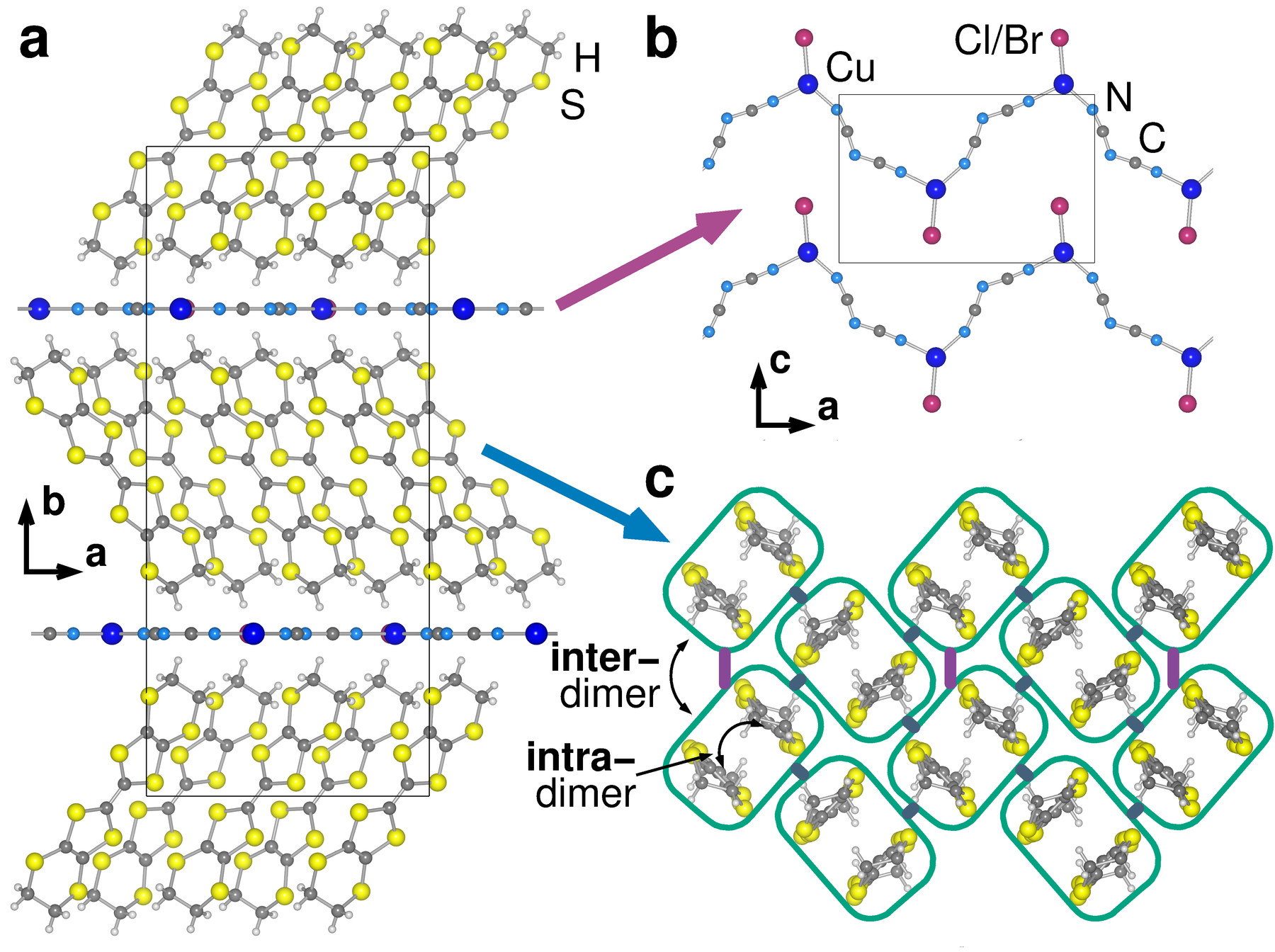}
\caption{\label{fig:structure} {\bf a} Structure of
  $\kappa$-(ET)$_2$Cu[N(CN)$_2$]Cl/Br seen in the $ab$ plane. {\bf b},
  {\bf c} Separate anion and cation layers projected in the $ac$
  plane. }
\end{figure*}

One of the most intensively debated open questions in condensed matter
physics is the emergence of exotic phases like spin liquid or
unconventional superconductivity in a Mott insulator upon changes in
temperature, doping or pressure.  In particular, an increasingly
prominent class of materials with an abundance of correlated phases
are the $\kappa$-based organic charge transfer salts containing the
molecules bis-(ethylenedithio)tetrathiafulvalene (BEDT-TTF, or shorter
ET).  In these $\kappa$-(ET)$_2X$ salts, electron donors (ET) and
electron acceptors ($X$) form alternating layers, with pairs of ET
molecules forming dimers (ET)$_2$ arranged in a triangular lattice
(see Figure~\ref{fig:structure}). For monovalent anions $X$ one
electron is transferred from each dimer (ET)$_2$ to each anion formula
unit so that the system is half-filled.  Band structure
calculations~\cite{Kandpal2009, Nakamura2009a} therefore predict the
dimer layers to be metallic. However, the experimentally observed
ground state depends on the choice of the anion: even for the example
of the isostructural compounds {\etcucl} (in short $\kappa$-Cl) and
{\etcubr} ($\kappa$-Br), the ground state can be as different as a
Mott insulator for $\kappa$-Cl and a Fermi liquid for $\kappa$-Br at
low temperatures and ambient pressure.~\cite{Note1} $\kappa$-Cl can be
driven through the insulator-to-metal transition (MIT) by the gradual
substitution of Cl for isovalent Br.  In the high-temperature regime
$\kappa$-Cl is a semiconductor with a gap of $E_g=
800$~K~\cite{Yasin2011} while $\kappa$-Br shows 'bad metal' behaviour
with strong scattering preventing coherent transport and suppressing
the Drude peak.

The fact that small chemical modifications lead to qualitative changes
in behaviour together with the possible importance of electronic
correlations in these materials make it clear that a realistic
description requires both (i) details of the band structure as well as
(ii) a proper treatment of electronic correlations. However, many-body
studies of the $\kappa$-(ET)$_2X$ salts have so far been limited to
minimal model calculations~\cite{Merino2000,Parcollet2004,Merino2008}
of the Hubbard or extended Hubbard Hamiltonian on an anisotropic
triangular lattice.~\cite{Powell2006} In this work, we go beyond model
calculations and present a combination of ab-initio density functional
theory calculations in the full potential linearised augmented plane
wave (FLAPW) framework~\cite{Blaha2001} combined with
DMFT~\cite{Aichhorn2009} (local density approximation plus dynamical
mean field theory, LDA+DMFT). Also, in order to perform these
calculations for organic systems, we propose a generalisation of
previously introduced Wannier projectors.~\cite{Aichhorn2009}

We show that this first principles approach provides insight into the
contributions to the optical conductivity at an unprecedented level;
(i) we find that interdimer and intradimer transitions are responsible
for two principal features resolved in optical conductivity
measurements at low temperatures, (ii) we unambiguously identify the
interdimer feature to be related to correlations and (iii) we are able
to unveil the origin of the anisotropic conductivity (light
polarisation-dependent).  These calculations go beyond previous
conjectures based on phenomenological and minimal model
assumptions.~\cite{Faltermeier2007,Dumm2009}

\begin{figure}[htb]
\includegraphics[width=0.4\textwidth]{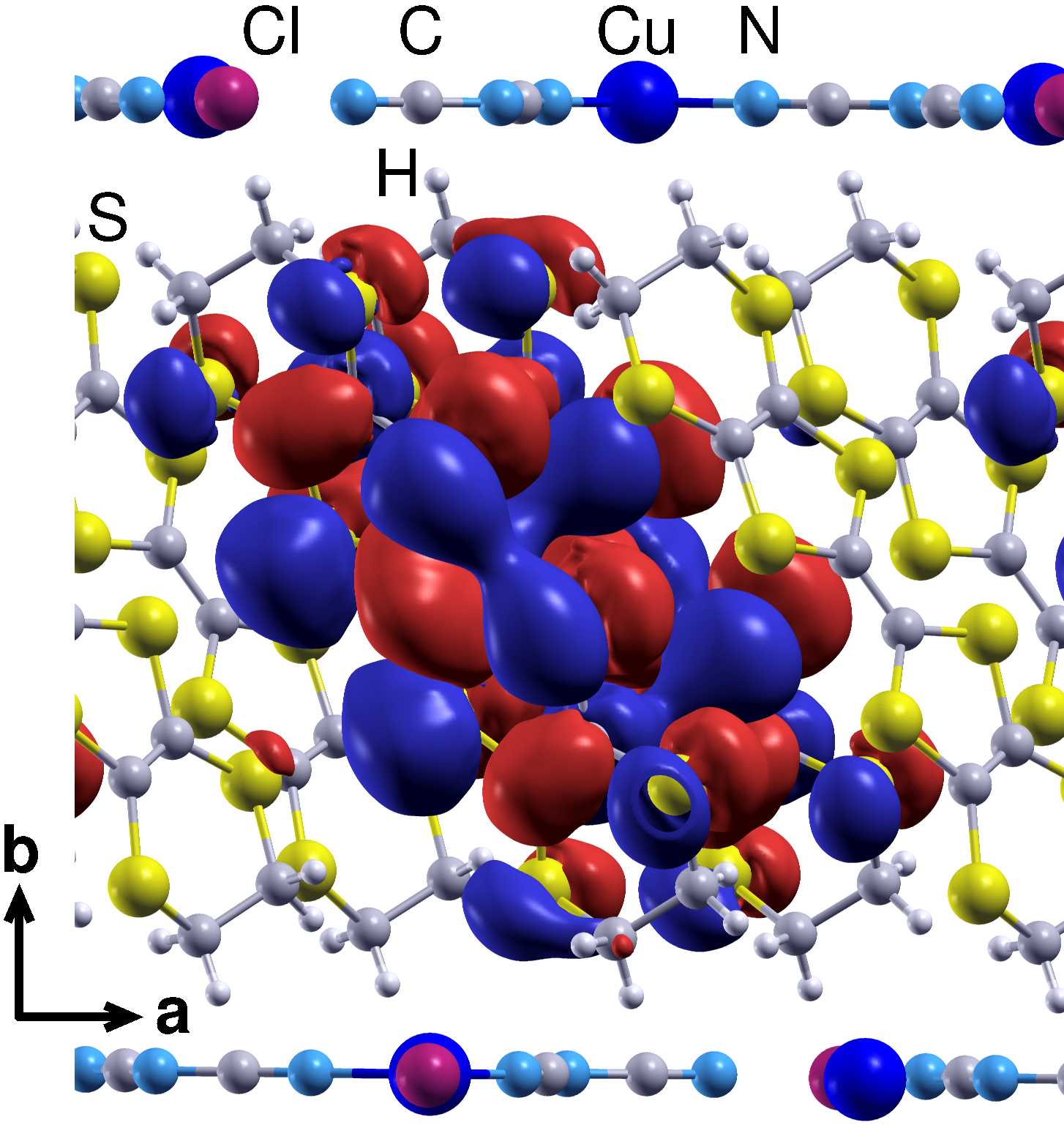}
\caption{\label{fig:wannier} Structure of {\etcucl} seen in the $ab$
  plane with a Wannier function corresponding to the bands crossing the
  Fermi level.}
\end{figure}

While recent LDA+DMFT calculations on organic molecular crystals
employed Wannier functions with a single atomic
character,~\cite{Weber2012} the electronic structure of organic charge
transfer salts requires the construction of localised Wannier
functions from the entire highest occupied molecular orbital (HOMO).
A fast and stable method to obtain Wannier functions consists in
projecting Bloch states onto pure atomic orbitals with subsequent
orthonormalization.~\cite{Ku2002,Anisimov2005,Aichhorn2009} In its
standard form, though, this method is not designed for molecular
orbitals.  In this work, we propose a scheme to construct molecular
Wannier functions using atomic orbitals as a starting point.  The key
element of this scheme is the diagonalization of the occupation matrix
written in the basis of atomic orbitals within the subspace of the
correlated bands.~\cite{notesupp} The real space representation of the
resulting dimer HOMO Wannier function of $\kappa$-Cl (based on the
crystal structure reported in Ref.~\onlinecite{Williams1990} with
space group $P\,nma$) is shown in Fig.~\ref{fig:wannier}.  Once the
Wannier functions are obtained out of the LDA cycle, we employ the
hybridisation expansion continuous-time quantum Monte Carlo
method~\cite{Werner2006} as implemented in the ALPS
code~\cite{Bauer2011,Gull2011} in order to solve the impurity problem
in the DMFT cycle. We used $2\times10^7$ Monte Carlo sweeps throughout
our calculations at an inverse temperature $\beta=40$~eV$^{-1}$,
corresponding to room temperature $T=300$~K.  Since the quantum Monte
Carlo algorithm operates on the imaginary frequency axis, the
calculation of dynamical quantities like spectral functions and
optical conductivity requires analytic continuation to the real
axis. We performed stochastic analytic continuation~\cite{Beach2004}
on the self energy for obtaining the spectral functions and directly
on the optical conductivity $\sigma(i\nu)$ for the calculation of
optical properties.~\cite{notesupp}

\begin{figure*}[htb]
\includegraphics[width=0.95\textwidth]{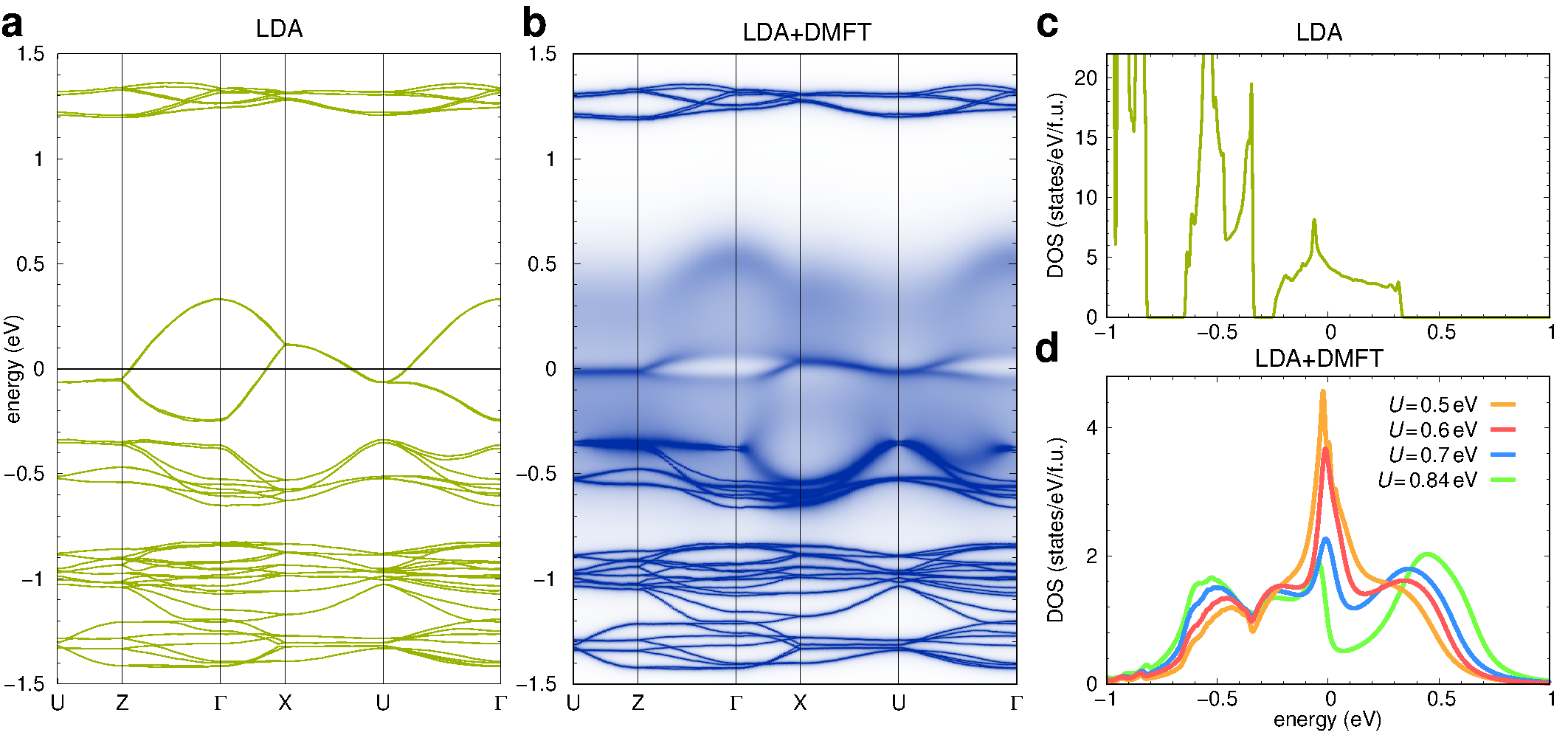}
\caption{\label{fig:Ak} {\bf a} LDA Bandstructure of {\etcucl}. {\bf
    b} LDA+DMFT momentum-resolved spectral function for $U=0.6$~eV.
  Note that the LDA+DMFT results are not sharp since the spectral
  function $A(k,\omega)$ is not a delta function. {\bf c} LDA density
  of states, and {\bf d} LDA+DMFT DOS ({\bf k} integrated spectral
  function, for the projected manifold) for different values of the
  interaction strength $U$.}
\end{figure*}

In Fig.~\ref{fig:Ak}~{\bf a}-{\bf b} we show the LDA+DMFT calculated
band structure~\cite{comment_unitcell} of {\brcl} in form of the
momentum-resolved spectral function for Hubbard $U=0.6$ eV
(Fig.~\ref{fig:Ak} {\bf b}) along with the LDA band energies
(\ref{fig:Ak} {\bf a}).  The LDA calculations were done for the
stoichiometric $\kappa$-Cl and $\kappa$-Br. We find only small changes
of the LDA bandstructure~\cite{Kandpal2009} at $E_{\rm F}$ between
both systems. The choice of $U=0.6$~eV in the LDA+DMFT calculations is
guided by two estimates: $U=0.85$~eV obtained for a similar but
arguably more strongly correlated compound
$\kappa$-(ET)$_2$Cu$_2$(CN)$_3$ from constrained random phase
approximation,~\cite{Nakamura2009a} and $U\approx 0.27$~eV extracted
from optical conductivity measurements~\cite{Faltermeier2007} and
model considerations.~\cite{Merino2008} The LDA+DMFT bands
(Fig.~\ref{fig:Ak} {\bf b}) show a strong renormalisation at $E_{\rm
  F}=0$ with respect to the LDA bands (Fig.~\ref{fig:Ak} {\bf
  a}). These bands originate from the interdimer hopping (see
Fig.~\ref{fig:structure} {\bf c}), in particular hopping between
dimers on the same layer; the interlayer hopping is very small, so
that the four bands are composed of two almost degenerate pairs of
bands.  The correlation in band space acts almost exclusively on these
bands, splitting them into renormalized excitations of (mass-enhanced)
quasiparticles (blue-scale colour map at $E_{\rm F}=0$) and a spectral
weight transfer to an upper and lower Hubbard band which manifests
itself as blurry background around -0.5~eV and 0.5~eV respectively.
On the other hand, the charge transfer between the ET molecules within
a dimer (intradimer, see Fig.~\ref{fig:Ak} {\bf c}) induces the
splitting between the bands into bonding (around $-0.4$~eV) and
antibonding bands (around the Fermi level).  This splitting is less
affected by correlations as can be observed by comparing
Fig.~\ref{fig:Ak} {\bf a} and {\bf b}.

Next, we investigate the optical properties of {\brcl} first with the
light polarisation {\Epc} along the linear chains in the triangular
lattice as measured experimentally.~\cite{Faltermeier2007,Dumm2009} At
room temperature, the authors of Ref.~\onlinecite{Faltermeier2007}
observed a broad mid-infrared absorption peak between 1600~cm$^{-1}$
and 4200~cm$^{-1}$, in agreement with previous optical studies on
$\kappa$-Cl and $\kappa$-Br (see Ref.~\onlinecite{Faltermeier2007} and
references therein).  At low temperature, a Drude peak evolves for the
compounds with high Br concentration which marks the onset of
metallicity at $x\approx 0.7$ whereas no Drude peak is visible for
lower Br content indicating an insulating state without coherent
quasiparticles.  Importantly, at low temperatures (T=90~K and below)
the broad mid-infrared peak (polarisation {\Epc}) splits into two
peaks in the pure Cl and low Br concentration compounds, which can be
fitted by two Lorentzians at $\approx\!2200$ cm$^{-1}$ and 3200
cm$^{-1}$; for high Br content this splitting is very weak but it is
still present. From this doping dependence, it was
suggested~\cite{Faltermeier2007} that the first peak is a
correlation-induced feature due to electron transitions between the
lower and upper Hubbard bands, while the second peak was assigned to
the intradimer charge transfer (see Fig. 8 of
Ref.~\onlinecite{Faltermeier2007}).  In contrast, for polarisation
{\Epa} the broad mid-infrared peak doesn't show any splitting at low
temperatures. This anisotropic behaviour of the optical conductivity
has remained unresolved up to now.

\begin{figure*}[htb]
\includegraphics[width=0.95\textwidth]{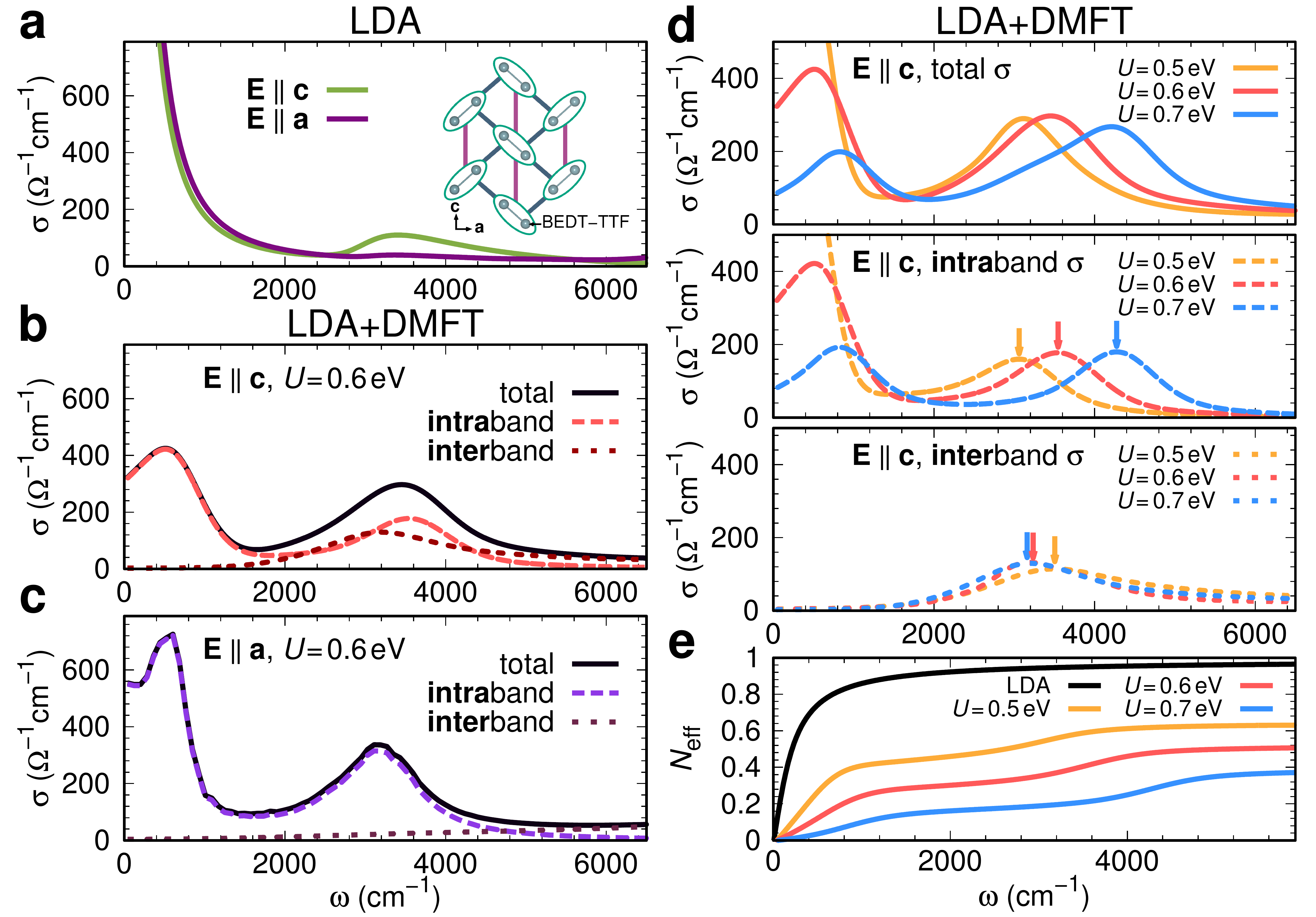}
\caption{\label{fig:sigma_U06}{\bf a} Calculated LDA optical
  conductivity for two directions of the light polarisation. {\bf b}
  and {\bf c} LDA+DMFT optical conductivity for $U=0.6$~eV for
  electrical field along {\bf a} and {\bf c}, respectively. {\bf d}
  LDA+DMFT optical conductivity for different values of the
  interaction strength $U$.  The arrows show the position of maxima
  for the intraband (middle panel) and the interband (lower panel)
  contributions. {\bf e} Optical conductivity sum rule for LDA and
  LDA+DMFT at different $U$ values.}
\end{figure*}

The method introduced in this work allows us to investigate the
microscopic origin of the observed spectra.  As the energy window of
our LDA+DMFT calculation contains both the correlated manifold at the
Fermi energy as well as uncorrelated bands away from $E_{\rm F}$, all
transitions can be inspected on equal footing.  Our calculations are
at T=300~K (continuous-time QMC) and capture the effects of
correlation contained in the T=300~K data and observed as pronounced
features in measurements below room temperature.  We denote the
transitions occurring at the Fermi level between (i) Hubbard bands and
(ii) Hubbard bands and quasiparticle peak (if present) as {\it
  intra}band contributions (i.e. {\it inter}dimer transitions) to the
optical conductivity.  All other transitions are termed {\it
  inter}band transitions; among others, these contain transitions
related to the {\it intra}dimer charge transfer.

In Fig.~\ref{fig:sigma_U06}~{\bf a}-{\bf c} we show the LDA and
LDA+DMFT ($U=0.6$~eV) calculated optical conductivity for polarisation
{\Epc} and {\Epa}.  We would like to note that while the LDA results
change minimally between $\kappa$-Cl and $\kappa$-Br, they are
extremely sensitive to the effects of correlation ($U$) as can be
observed in the density of states (Fig.~\ref{fig:Ak}~{\bf c}-{\bf d})
and in the optical conductivity (Fig.~\ref{fig:sigma_U06}~{\bf d}).
For $U >0.84$~eV a gap opens and the Drude peak disappears.  This
corresponds to the $\kappa$-Cl system although with a somewhat
overestimated $U$ value (this is a well-known limitation of
single-site DMFT in low dimensions~\cite{Merino2008}). Smaller $U$
values feature the conductivity behaviour of {\brcl} at larger values
of $x$.  For $U=0.6$~eV we observe a remnant of the Drude peak as it
is the case in {\brcl} at moderate $x$.

For {\Epc} the calculated LDA+DMFT total optical conductivity
(Fig.~\ref{fig:sigma_U06}~{\bf b}) features one dominant low-frequency
peak at approximately 3450~cm$^{-1}$, {\ie}, close to the experimental
peak position (we do not consider vibronic
modes).~\cite{Faltermeier2007} The position of this peak is roughly
centred at the same position as the LDA results
(Fig.~\ref{fig:sigma_U06} {\bf a}), but it is strongly enhanced in
spectral weight, in accordance with experiment.~\cite{Faltermeier2007}
While our calculations are at room temperature, they already capture
the effects of correlation resolved in measurements at low T. We can
decompose our data into intra- and interband contributions.
$\sigma_{\rm interband}$ contains contributions of transitions between
almost uncorrelated bands and roughly coincides with the LDA results
in position and spectral weight. This contribution shows a peak at
$\omega\!\approx\! 3400 \mbox{cm}^{-1}$ that corresponds to {\it
  intra}dimer transitions (Fig.~\ref{fig:sigma_U06} {\bf b} dotted
line).  In contrast, the intraband contribution $\sigma_{\rm
  intraband}$ is centred at $\omega\!\approx\! 3550
\mbox{cm}^{-1}\!\approx\! 0.75 \, U$ and it corresponds to the
intraband Hubbard transitions ({\it inter}dimer contribution,
Fig.~\ref{fig:sigma_U06} {\bf b} dashed line).  Once thermal
fluctuations are suppressed by lowering temperature, the two low
frequency contributions ($3400 \mbox{cm}^{-1}$ and ($3550
\mbox{cm}^{-1}$) --that come naturally out of our calculations-- are
experimentally observed as two distinguishable peaks, a "dimer" peak
and a "Hubbard" peak respectively.~\cite{Faltermeier2007}

In order to analyse the nature of the two low-frequency intraband and
interband absorption peaks, Fig.~\ref{fig:sigma_U06}~{\bf d} shows
their evolution with $U$. The intraband contribution (interdimer
transitions) to the conductivity shifts in frequency proportional to
$U$, with the peak position consistently corresponding to
\mbox{$\approx 0.75\,U$} while the interband absorption peak at low
frequencies (intradimer transitions) is largely insensitive to $U$.
This analysis demonstrates the correlated nature of $\sigma_{\rm
  intraband}$ and uncorrelated nature of $\sigma_{\rm interband}$.

Quantitatively, the suppression of the Drude peak as a function of $U$
and the redistribution of the intraband spectral weight is presented
in Fig.~\ref{fig:sigma_U06}~{\bf e} where we plot the integrated
spectral weight $\int_0^\omega \sigma_{\rm intraband}(\omega')
d\omega'$ representing the effective number of charge carriers $N_{\rm
  eff}$. In this representation, the number of charge carriers in LDA
by definition equals the number of conduction electrons, {\ie} one,
and all the weight is concentrated in the (infinitesimally narrow)
coherent Drude peak which is only broadened by temperature. Upon
inclusion of correlations, the kinetic energy of the electrons is
diminished, which corresponds to a mass enhancement (in Fermi liquid
theory) or a reduction of the number of effective charge carriers as
we observe.

Summarising the above analysis, we have shown from first principles
that the two finite-frequency peaks resolved in the experimental
optical conductivity for {\Epc} of {\brcl} at low temperatures
originate, respectively, from correlation-induced intraband
(interdimer) contributions scaling with $U$, and interband
(intradimer) transitions which are unaffected by correlations.  The
fact that DMFT overestimates the critical $U$ of the triangular
lattice in two dimensions forces us to describe these systems at
T=300~K with a somewhat high interaction strength of
\mbox{$U=0.6$~eV}.

We focus now on the optical conductivity {\Epa}.  The LDA optical
conductivity is completely featureless for \mbox{$\omega\approx 3400\:
  {\rm cm}^{-1}$} (Fig.~\ref{fig:sigma_U06} {\bf a}).  Inclusion of
correlation effects (Fig.~\ref{fig:sigma_U06} {\bf c}) shows the
appearance of a peak at this frequencies. This peak is therefore only
a consequence of transitions between the lower Hubbard
band{\slash}quasiparticle peak and the upper Hubbard band (interdimer
contribution) and at low temperatures no splitting of the peak is
therefore to be expected, as observed
experimentally~\cite{Faltermeier2007}, in contrast to the {\Epc} case.
Finally we would like to note that while calculations were performed
at T=300 K, important information was obtained for the features
observed at lower temperatures.  Accurate calculations of the optical
conductivity at lower temperatures requires the use of alternative
impurity solvers~\cite{Deng2013} and is beyond the scope of the
present study.

In conclusion, we presented the first LDA+DMFT study on the spectral
and optical properties of the organic charge transfer salts {\brcl}.
Our results provide an {\it ab initio}-based theoretical evidence for
the double-natured origin of the infrared peak in the optical
conductivity of this system for {\Epc} as well as for the
single-natured origin of the infrared peak for {\Epa}.  We could
identify intraband transitions within the correlated manifold and
interband transitions due to charge-transfer within an ET dimer.  The
proposed projection method for constructing non-atom-centred Wannier
functions in the FLAPW framework is computationally efficient and can
be applied to a great variety of correlated organic as well as
inorganic systems with (quasi-)molecular orbitals. This opens the
possibility of investigating correlation effects in complex organic
systems from first principles.

{\it Acknowledgements.-} We acknowledge useful discussions with
M. Dressel. We gratefully acknowledge financial support from the
Deutsche Forschungsgemeinschaft through TR49 and FOR1346 and the
allotment of computer time by CSC-Frankfurt and LOEWE-CSC.


\begin{thebibliography}{99}
\bibitem{comment_unitcell}
Note that in the crystallographic unit cell, there is a manifold of
four bands around $E_{\rm F}$ as the crystallographic unit cell
contains two organic layers with a total of four dimers due to the
presence of the anion. The dimers are equivalent, though, so that only
one dimer HOMO has to be considered in our single-site DMFT; the other
three orbitals are related by symmetry.
\end{thebibliography}
\end{document}